\documentclass[letter]{aa}
\usepackage{txfonts}
\usepackage{graphicx}
\usepackage{natbib}
\begin{document}

\title{Bayesian re-analysis of the radial velocities of Gliese 581}
\subtitle{Evidence in favour of only four planetary companions}
\author{Mikko Tuomi\inst{1,2}\thanks{\email{mikko.tuomi@utu.fi}}}

\institute{University of Hertfordshire, Centre for Astrophysics Research, Science and Technology Research Institute, College Lane, AL10 9AB, Hatfield, UK \and University of Turku, Tuorla Observatory, Department of Physics and Astronomy, V\"ais\"al\"antie 20, FI-21500, Piikki\"o, Finland}

\date{Received xx.xx.xxxx / Accepted xx.xx.xxxx}

\abstract{}
{The Gliese 581 planetary system has received attention because it has been proposed to host a low-mass planet in its habitable zone. We re-analyse the radial velocity measurements reported to contain six planetary signals to see whether these conclusions remain valid when the analyses are made using Bayesian tools instead of the common periodogram analyses.}
{We analyse the combined radial velocity data set obtained using the HARPS and HIRES spectrographs using posterior sampling techniques and computation of the posterior probabilities of models with differing numbers of Keplerian signals. We do not fix the orbital eccentricities and stellar jitter to certain values but treat these as free parameters of our statistical models. Hence, we can take the uncertainties of these parameters into account when assessing the number of planetary signals present in the data, the point estimates of all of the model parameters, and the uncertainties of these parameters.}
{We conclude that based on the Bayesian model probabilities and the nature of the posterior densities of the different models, there is evidence in favour of four planets orbiting GJ 581. The HARPS and HIRES data do not imply the conclusion that there are two additional companions orbiting GJ 581. We also revise the orbital parameters of the four companions in the system. Especially, according to our results, the eccentricities of all the companions in the system are consistent with zero.}
{}


\maketitle


\section{Introduction}

The nearby M dwarf Gliese 581 has received plenty of attention during recent years. In 2005 a Neptune-mass planet candidate was found in its orbit using the HARPS spectrograph \citep{bonfils2005}. Two years later it was reported to be a host to two additional super-Earths \citep{udry2007}. In 2009 an new Earth-mass planet with a minimum mass of 1.9 M$_{\oplus}$ was found in its orbit, making it a system with four planetary companions of relatively low mass \citep{mayor2009}.

Ever since the discovery of the first companion orbiting GJ 581, the star has been a target of intensive radial velocity (RV) surveys because few M dwarf stars are known to be hosts to planetary systems despite the fact that they are numerous even in the Solar neighbourhood. As a result, in 2010, GJ 581 was reported to have two more companions of planetary mass with GJ 581 g, a 3.1 M$_{\oplus}$ planet, in the habitable zone of the star \citep{vogt2010}.

The discovery of this 6-planet system was made by analysing two high-precision RV data sets made using the HARPS spectrograph \citep{mayor2009} and the HIRES spectrograph \citep{vogt2010}. The combined data set consists of 241 RV measurements. In \citet{vogt2010}, the Keplerian signals of the six planets were discovered by studying the periodogram of the combined timeseries and by fitting the orbital parameters of the proposed companions to the data.

The purpose of this letter is to see whether Bayesian data analysis gives consistent results with those reported by \citet{vogt2010}. Our major concern is, that by fixing eccentricities to zero the uncertainties of these eccentricities and their effect on the detectability of planetary signals were not taken into account by \citet{vogt2010}. They did let the eccentricities float freely and concluded that this did not produce any significan improvement to the fit. However, they did not discuss whether the uncertainty about the eccentricities could have an effect on the probabilities of finding the Keplerian signals in the data in the first place. Also, in \citet{vogt2010}, the stellar jitter was estimated to have a value of 1.4 ms$^{-1}$ -- simply because it yielded a reduced $\chi^{2}$ value of unity. Our second concern is that the uncertainty of the jitter was not taken into account either in the analyses. If its value was under- or overestimated, it could have a significant effect on the detectability of the planetary signals as demonstrated by e.g. \citet{ford2006,gregory2007a,gregory2007b,tuomi2009}.

In this letter, we reanalyse the combined RV data set of GJ 581 using Bayesian tools -- posterior samplings and model probabilities. First, we sample the parameter spaces of Keplerian models with k planetary companions by letting $k = 0, ..., 6$. Second, we calculate the Bayesian probabilities for each of these models $\dot{z}_{k}$. Finally, we compare our results with those of \citet{vogt2010} to see whether their periodogram analyses and fitting algorithms and Bayesian ones do yield similar results in this case.

\section{Modelling and Bayesian model comparison}

\subsection{Statistical model and posterior samplings}

Following \citet{tuomi2009}, we assume that the planets do not interact with one another in the timescale of the measurements and model the superposition of $k$ Keplerian signals simply by summing their effect on the RV. Consequently, there are five parameters describing the signature of an individual planet: RV amplitude ($K$), orbital period ($P$), orbital eccentricity ($e$), longitude of pericentre ($\omega$), and the mean anomaly ($M_{0}$). As suggested by \citet{ford2006}, to improve the efficiency of the sampling of the parameter space, we use the logarithm of the period in the samplings because it is a scale-invariant parameter.

Our statistical model consists of the sum of Keplerian signals and two sources of uncertainty. These sources are the instrument noise and the noise caused by the stellar surface -- the stellar jitter. We model these as independent random variables with Gaussian density and zero mean. We assume that the standard deviation of the instrument noise is known and use the values reported in \citet{mayor2009,vogt2010}. However, we adopt the standard deviation of the stellar jitter as a free parameter of our statistical model and denote it as $\sigma_{J}$. The statistical model can be written as
\begin{equation}\label{rv_model}
  r_{i,l} = \dot{z}_{k}(t_{i}) + \gamma_{l} + \epsilon_{i} + \epsilon_{J} ,
\end{equation}
where $r_{i,l}$ is the measurement at time $t_{i}$ made using telescope-instrument combination $l$, $\dot{z}_{k}$ represents the $k$ Keplerian signals, parameter $\gamma_{l}$ is the reference velocity, and $\epsilon_{i}$ and $\epsilon_{J}$ are Gaussian random variables describing the instrument noise, as reported by the observers, and the stellar jitter, respectively. As a result, there are $5k + 3$ free parameters in the parameter vectors $\theta_{k}$ of our models -- five parameters for each planet, jitter magnitude, and the reference velocities of the HARPS and HIRES measurements.

We sample the parameter spaces of the different models using the adaptive Metropolis algorithm of \citet{haario2001}. This sampling algorithm is similar to the famous Metropolis-Hastings algorithm \citep{metropolis1953,hastings1970} but it adapts to the information gathered during the first $n - 1$ samples from the parameter posterior density by approximating this sample as a multivariate Gaussian density. The $n$th sample is then drawn by using this multivariate Gaussian as a proposal density with the $n - 1$th value as a mean. While only an approximation, this algorithm converges to the posterior relatively rapidly -- apparently even in the case of multimodal posterior sampled in this letter.

When calculating the posterior densities for the parameters representing semimajor axes and RV masses of the planets, we took into account the uncertainty of the stellar mass. This mass is estimated to be $0.31 \pm 0.02$ M$_{\odot}$ for GJ 581 \citep{delfosse2000}. We sampled the densities of the semi-major axes and RV masses by drawing random numbers from the estimated density of the stellar mass that had a mean of 0.31 M$_{\odot}$ and a standard deviation of 0.02 M$_{\odot}$. This enabled us to calculate more reliable estimates for the uncertainties of the semimajor axes and RV masses.

\subsection{Model comparisons}

We compare the models with differing number of planetary companions using the Bayesian model probabilities. The probability of the $k$th model is defined as
\begin{equation}\label{model_probability}
  P(\dot{z}_{k} | r) = \frac{P(r | \dot{z}_{k}) P(\dot{z}_{k})}{\sum_{j=0}^{p} P(r | \dot{z}_{j})P(\dot{z}_{j})} ,
\end{equation}
where the marginal integral is
\begin{equation}\label{marginal}
  P(r | \dot{z}_{k}) = \int f(r | \theta_{k}, \dot{z}_{k}) p(\theta_{k} | \dot{z}_{k}) d \theta_{k} ,
\end{equation}
$f(r | \theta_{k}, \dot{z}_{k})$ is the likelihood function, and $p(\theta_{k} | \dot{z}_{k})$ the prior density of the model parameters. In addition, $P(\dot{z}_{k})$ is the prior probability of the $k$th model and $p$ is the greatest number of planetary signals in our analyses.

We interpret the probabilities of Eq. (\ref{model_probability}) as proper probabilities and require that the probability
of confidently finding a $k$th planetary signal requires that $P(\dot{z}_{k}) \gg P(\dot{z}_{k-1})$. In practice, according to \citet{jeffreys1961}, we require that the probability of finding $k$ signals is at least 150 times greater than that of finding $k - 1$ signals to be able to claim confidently that there are $k$ planets orbiting the target star. In addition, we require that the probability density of the $k$th planet has a unique maximum that can be interpreted as a Keplerian signal and is not likely to be caused by gaps in the data or by pure noise.

Since the integral in Eq. (\ref{marginal}) cannot be calculated analytically, we calculate its value from the sample from the posterior density using the technique discussed in \citet{chib2001}.

\section{Results}

We modelled the system by making two different assumptions about the orbital eccentricities of the planets. First, we treated the eccentricities as free parameters of the model, letting them vary freely in the samplings of the parameter spaces. Second, following the analysis of \citet{vogt2010}, we tested a case where the eccentricities were only allowed to have low values -- values below 0.2 -- for the sake of dynamical stability of the system.

\subsection{Eccentricities as free parameters}

When adopting the orbital eccentricities of the $k$ planets in the system as free parameters of the model, we receive the model probabilities ($P_{A}$) in Table \ref{probabilities}. The probabilities of both model sets $P_{A}$ and $P_{B}$, with different assumptions on the eccentricity, are scaled to unity according to Eq. (\ref{model_probability}).

In Table \ref{probabilities}, we present the probabilities of the different models with $k = 0, ..., 6$ planetary companions. We divide the model with $k=5$ into two different solutions. In these solutions, 5a and 5b are used to denote the 5-companion models including the four planets reported by \citet{mayor2009} and a fifth Keplerian signal corresponding to one or other of the two signals proposed by \citet{vogt2010} with periods of roughly 37 and 433 days, respectively. The 6 planet model includes all six planets reported by \citep{vogt2010}. Clearly, according to the probabilities, it cannot be concluded that there are the signals of six planetary companions in the data set. Instead, allowing the values of the eccentricities to be determined freely by the data leads to a conclusion that there are clear signals of at least four companions in the data but the 4-companion model has such high probability that it cannot be ruled out confidently enough to claim that there are more than four Keplerian signals in the data. Hence, taking into account the uncertainties of the orbital and other parameters of the models leads to results that contradict with those of \citet{vogt2010}.

\begin{table}
\center
\caption{Bayesian model probabilities of $k$ planet models with free eccentricity ($P_{A}$) and eccentricity limited to values below 0.2 ($P_{B}$). The models 5a and 5b correspond to the solutions with the orbital period of GJ 581 f roughly at 37 and 433 days, respectively.\label{probabilities}}
\begin{tabular}{lcc}
\hline \hline
  $k$ & $P_{A}$ & $P_{B}$ \\
\hline
  0 & $< 10^{-128}$ & $< 10^{-126}$ \\
  1 & $< 10^{-33}$ & $< 10^{-31}$ \\
  2 & $< 10^{-13}$ & $< 10^{-12}$ \\
  3 & $< 10^{-10}$ & $< 10^{-8}$ \\
  4 & 0.01 & 0.01 \\
  5a & $< 10^{-3}$ & $< 10^{-2}$ \\
  5b & 0.98 & 0.34 \\
  6 & 0.01 & 0.64 \\
\hline \hline
\end{tabular}
\end{table}

The 5-planet solution with eccentricities as free parameters consists broadly of two clear probability maxima for the 5th planet (the roughly 37 and 433 day periodicities) but we cannot conclude that the corresponding signals are real as opposed to artefacts of the data. Interestingly, the 433 day periodicity proposed by \citet{vogt2010} has a greater probability than the 37 day periodicity but the former appears to consist of two closely spaced maxima in the periodicity space. There is one additional maxima in the vicinity of this period (at roughly 465 days), as can be seen in Fig. \ref{period}.

\begin{figure}
\begin{center}
\includegraphics[angle=270, width=0.4\textwidth]{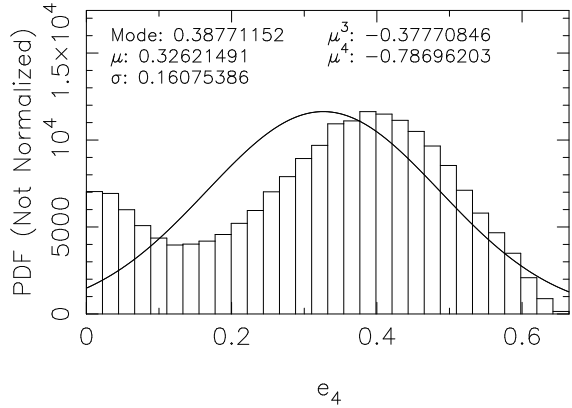}
\end{center}
\caption{The distribution of the orbital eccentricity of GJ 581 d from the 4-companion solution. A Gaussian curve with the same mean and variance is shown for comparison together with the mode, mean ($\mu$), standard deviation ($\sigma$), skewness ($\mu^{3}$), and kurtosis ($\mu^{4}$) of the distribution.}\label{eccentricity}
\end{figure}

\begin{figure}
\begin{center}
\includegraphics[angle=270, width=0.4\textwidth]{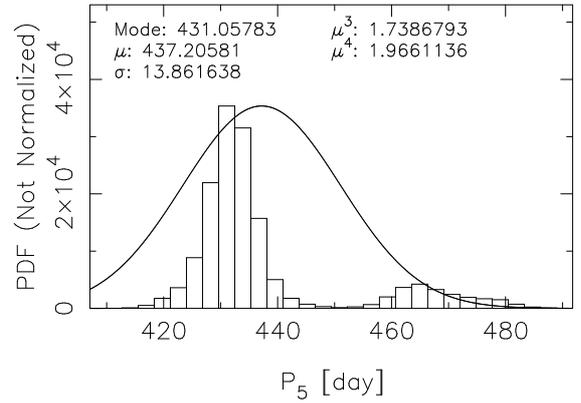}
\end{center}
\caption{Same as Fig \ref{eccentricity} but for the period of GJ 581 f in the 5-companion solution.}\label{period}
\end{figure}

If the orbital stability of the system is not considered, there is evidence in the data in favour of only four planetary companions. Curiously, we receive an interesting probability distribution for the eccentricity of planet GJ 581 d. This distribution is shown in Fig. \ref{eccentricity} and it supports the conclusion of \citet{mayor2009}, who claimed that this companion has an eccentricity of $0.38 \pm 0.09$. However, there is another maximum in this distribution close to zero, which makes the eccentricity of GJ 581 d consistent with zero. The eccentricities of the other companions were also consistent with zero, yet that of GJ 581 c peaked at 0.1 -- a value consistent with the results of \citet{mayor2009}.

According to our results, the point estimates of the orbital parameters and especially their uncertainty estimates need to be revised. The revised parameters and their 99\% Bayesian credibility sets ($\mathcal{D}_{0.99}$), as defined in e.g. \citet{tuomi2009}, are shown in Table \ref{parameters}.

\begin{table*}
\center
\caption{The four-planet solution of GJ 581 RV's. Maximum \emph{a posteriori} estimates of the parameters and their $\mathcal{D}_{0.99}$ sets.\label{parameters}}
\begin{tabular}{lcccc}
\hline \hline
Parameter & GJ 581 b & GJ 581 c & GJ 581 d & GJ 581 e \\
\hline
$P$ [days] & 5.36849 [5.36810, 5.36888] & 12.916 [12.909, 12.922] & 66.85 [65.85, 67.76] & 3.1488 [3.1479, 3.1510] \\
$e$ & 0.007 [0, 0.051] & 0.09 [0, 0.29] & 0.39 [0, 0.67] & 0.08 [0, 0.43] \\
$K$ [ms$^{-1}$] & 12.50 [11.84, 13.11] & 3.36 [2.71, 3.95] & 1.58 [0.86, 2.22] & 1.73 [1.06, 2.33] \\
$\omega$ [rad] & 2.4 [0, 2$\pi$] & 2.4 [0, 2$\pi$] & 5.3 [0, 2$\pi$] & 1.8 [0, 2$\pi$] \\
$M_{0}$ [rad] & 1.4 [0, 2$\pi$] & 3.1 [0, 2$\pi$] & 4.9 [0, 2$\pi$] & 1.6 [0, 2$\pi$] \\
$m_{p} \sin i$ [M$_{\oplus}$] & 15.70 [13.50, 17.70] & 5.53 [4.40, 6.93] & 4.51 [2.59, 6.69] & 1.81 [1.07, 2.55] \\
$a$ [AU] & 0.0405 [0.0380, 0.0430] & 0.0729 [0.0683, 0.0775] & 0.218 [0.203, 0.232] & 0.0283 [0.0267, 0.0303] \\
\hline
$\gamma_{1}$ [ms$^{-1}$] (HARPS) & 1.13 [0.48, 1.78] \\
$\gamma_{2}$ [ms$^{-1}$] (HIRES) & -0.35 [-0.95, 0.19] \\
$\sigma_{J}$ [ms$^{-1}$] & 1.89 [1.51, 2.36] \\
\hline \hline
\end{tabular}
\end{table*}

\subsection{Low eccentricities}

We further assumed that the orbital eccentricities had values lower than 0.2 and repeated the analyses in the previous subsection. The results of these analyses show that the five- and six-planet models do not have great enough posterior probabilities to be able to claim that there are more than four planetary companions orbiting GJ 581 (Table \ref{probabilities}). These results show that the six-companion solution proposed by \citet{vogt2010} cannot be considered to imply the existence of six planets in the system because the four-companion model cannot be shown to be an insufficient description of the data confidently enough. Instead, the solution of \citet{mayor2009} remains the most convincing solution even with the combined HARPS and HIRES dataset. Also, the updated parameter and uncertainty estimates of this solution are those presented in Table \ref{parameters}.

Regarding the existence of the proposed companion in the habitable zone of GJ 581 with an orbital period of roughly 37 days, the posterior probabilities in Table \ref{probabilities} imply that there is no evidence in favour of the existence of this companion. It is more probable that there is a companion corresponding to the periodicity of 433 days but even this periodicity appears to not be probable enough and also rather poorly constrained, as shown in Fig. \ref{period}.

We also note, that limiting the eccentricities of the companions to values lower than 0.2 favours the six-companion interpretation presented in \citet{vogt2010}. The reason is that the eccentricity of GJ 581 d likely differs from zero and the resulting solution where eccentricities are not limited describes the data much better than the limited case.

Based on the data alone, the freely varying eccentricity is a much more likely scenario than the limited one with roughly 25 times greater probability for the four-companion model and more than 100 times greater probability for the five-companion model. However, they are almost equal for the six-companion model, which means that the hypothetical six-companion model favours negligible eccentricities.

\subsection{Stellar jitter}

For the 4-companion solution, our estimate of stellar jitter of 1.89 [1.51, 2.36] ms$^{-1}$ appears to differ significantly from the value reported by \citet{mayor2009} of 1.2 ms$^{-1}$. First, we note that the jitter does not only correspond to the noise caused by the stellar surface, but it contains all the excess variations in the data not explained by the Keplerian model or the instrument noise. With measurements from two telescope-instrument combinations, any small differences in the calibration of the telescopes and instruments may cause systematic differences to the measurements and lead to increased values for the jitter. Also, undetected planets may increase the jitter between our results and those of \citet{mayor2009} because the observational timeline is longer in the combined dataset analysed here.

Second, our lower limit of the jitter, or more accurately excess noise, in the combined dataset is 1.51 ms$^{-1}$ -- a value reasonably close to the estimate of \citet{mayor2009}. The fact that we used posterior sampling technique, and took into account the uncertainties of all the parameters, can result in a greater estimate for the jitter magnitude because the common $\chi^{2}$ fitting, where jitter is not a free parameter, yields the lowest possible value for the jitter -- fitting only the orbital parameters and reference velocities essentially corresponds to minimising the excess noise, not finding the most probable solution.

\subsection{Orbital stability}

We studied the orbital stability of out four-planet solution (Table \ref{parameters}) briefly by using numerical integrations of the planetary orbits. We used the Bulirsch-Stoer algorithm \citep{bulirsch1966} because it is a relatively fast and reliable algorithm for studying the dynamics of planetary systems. To assess the instability of the system, we used the concept of Lagrange stability, in which the orbital parameters remain inside some bounded subspace of the parameter space for stable systems. Therefore, we considered the system not stable if there was orbital crossing, collision between the planets, accretion of a planet by the star, or if a planet escaped from the system with $a$ exceeding 100 AU.

We drew a sample of 50 parameter vectors from the parameter posterior density by weighting the sample towards the high-eccentricity orbits -- the orbital configurations most likely to be unstable -- allowed by our solution. We integrated the orbits for 10 000 years for each parameter vector because during that period of time even the outer planet would have completed more than 50 000 orbits. We checked whether the semi-major axes or orbital ecentricities evolved significantly during these integrations.

In our the numerical integrations of the planetary orbits, the semi-major axes and eccentricities remained bounded in all but ten integrations. The semi-major axes remained almost constant and the orbital ecentricities of the three inner planets librated roughly between 0 and 0.2, whereas the eccentricity of the outer planet librated only slightly around its initial value. Also, despite the possible moderate eccentricity of the outer planet of even more than 0.5, it did not appear to have a significant effect on the orbits of the three inner companions that would have resulted in orbital crossings. In ten cases, the initial eccentricities of planet c and e were greater than 0.2, where their probability densities have already become very low with respect to the MAP values. These eccentricities resulted in orbital crossings and it can be therefore concluded that these planets are likely to have eccentricities lower than 0.2. This is a consequence of the tight packing of the three innermost planets in the system.

With these constraints, our sample from the posterior appeared to correspond to stable orbital configurations. Therefore, though having analysed the stability only briefly, we conclude that our revised solution likely corresponds to a stable system.

\section{Conclusions and discussion}

According to our analyses of the combined RV data for Gliese 581 from HARPS and HIRES spectrographs, it cannot be concluded that there are six planetary companions orbiting Gliese 581. We find that there are confidently four keplerian signals in the combined data set and that there is no evidence in favour of the existence of a low-mass planet in the habitable zone on Gliese 581 with an orbital period of 37 days. Therefore, we conclude that our four-companion solution is consistent with the results of \citet{mayor2009}, though the uncertainties of the orbital parameters and RV masses require revision. We also conclude that the interpretation of \citet{vogt2010}, who claimed that there are as many as six planets orbiting Gliese 581, appears to have been not supported by the data strongly enough. There is a periodicity of roughly 433 days in the combined data but it is not probable enough to claim that it is a real signal as opposed to an artefact of the data spacing or pure noise.

We do realise that while the combined dataset provides evidence in favour of the existence of only four companions orbiting GJ 581, it cannot be claimed that the proposed solution of \citet{vogt2010} is not real -- it is fairly possible that the proposed companions f and g do exist in the system. However, the current data do not imply their existence in a statitically significant way, and the solution of the combined dataset remains that in Table \ref{parameters}.

The most likely reasons for the fact that we could not verify the results of \citet{vogt2010} are that we i) treated the orbital eccentricities of the planets as free parameters throughout the analyses; ii) adopted the magnitude of the stellar RV jitter as a free parameter of our statistical model instead of fixing its value to some \emph{a priori} determined value or minimising it; and iii) calculated the posterior probabilities of the different models that take into account the Occamian principle of parsimony in a consistent manner -- i.e. penalise the models more strongly the more free parameters they contain. We note that our estimate for the stellar jitter of 1.89 ms$^{-1}$ contradicts the value obtained by \citet{vogt2010} of 1.4 ms$^{-1}$ but is very close that estimated from the data of \citet{wright2005} of 1.9 ms$^{-1}$.

According to the dynamical studies of our solution (Table \ref{parameters}), our revised solution is likely stable -- a result consistent with the analyses of \citet{mayor2009}. However, the availability of a sample from the posterior density of the orbital parameters allows a more detailed dynamical study of the Gliese 581 system. Such a study could help ruling out some subsets of the parameter space that appear to have a high probability based on the data alone and could not be shown to correspond to unstable configurations in this study because of the relatively low integration times.

Additional high-precision measurements of the low-mass target Gliese 581 are needed to assess the number of super-Earths in its orbit. Also, the first low-mass planet confidently orbiting its host-star within the limits of the stellar habitable zone remains to be confirmed with confidence.

\begin{acknowledgements}
The author was supported by RoPACS (ROcky Planets Around Cool Stars), a Marie Curie Initial Training Network funded by the European Commission's Seventh Framework Programme.
\end{acknowledgements}


\end{document}